\documentclass[prl,twocolumn,floatfix,superscriptaddress,amsmath,citeautoscript,aps,longbibliography]{revtex4-2}
\pdfoutput=1
\usepackage[T1]{fontenc}\usepackage[latin1]{inputenc}
\usepackage{dcolumn,bm,graphicx,color,booktabs,microtype,afterpage} \graphicspath{{Figures/}}
\usepackage[charter,greekuppercase=italicized]{mathdesign}
\renewcommand{\figurename}{Fig.}
\renewcommand{\tablename}{Table}
\makeatletter\renewcommand{\fnum@figure}[1]{\figurename~\thefigure.}\makeatother
\makeatletter\renewcommand{\fnum@table}[1]{\tablename~\thetable.}\makeatother
\newcount\hh \newcount\mm
\hh=\time \divide\hh by 60
\mm=\hh \multiply\mm by 60 \mm=-\mm
\advance\mm by \time
\def\now{\number\hh:\ifnum\mm<10{}0\fi\number\mm}
\usepackage[colorlinks,plainpages=false,linkcolor=blue,urlcolor=blue,citecolor=blue,pdfpagemode=UseNone,pdfstartview=FitBH]{hyperref}

\newcommand{\mfs}{Mn$_{0.9}$Fe$_{0.1}$Si}

\begin{document}

\title{Anisotropy of spin waves in the field-polarized phase of Fe-doped MnSi}

\author{I.~N.~Khoroshiy}
\affiliation{Kirensky Institute of Physics, Federal Research Center, KSC, Siberian Branch Russian Academy of Sciences, 660036 Krasnoyarsk, Russia}
\affiliation{Department of Solid State Physics and Nanotechnology, Institute of Engineering Physics and Radioelectronics, Siberian Federal University, Krasnoyarsk 660041, Russia}

\author{A.~Podlesnyak}
\affiliation{Neutron Scattering Division, Oak Ridge National Laboratory, Oak Ridge, Tennessee 37831, USA}

\author{D.~Menzel}
\affiliation{Institut f\"ur Physik der Kondensierten Materie, Technische Universit\"at Braunschweig, D-38106 Braunschweig, Germany}

\author{M.~C.~Rahn}
\affiliation{Experimental Physics VI, Center for Electronic Correlations and Magnetism, University of Augsburg, 86159 Augsburg, Germany}

\author{D.~S.~Inosov}
\affiliation{Institut f{\"u}r Festk{\"o}rper- und Materialphysik, Technische Universit{\"a}t Dresden, D-01069 Dresden, Germany}
\affiliation{W\"urzburg-Dresden Cluster of Excellence on Complexity and Topology in Quantum Matter\,---\,\textit{ct.qmat}, TU Dresden}

\author{A.~S.~Sukhanov}
\affiliation{Experimental Physics VI, Center for Electronic Correlations and Magnetism, University of Augsburg, 86159 Augsburg, Germany}

\author{S.~E.~Nikitin}
\affiliation{Kirensky Institute of Physics, Federal Research Center, KSC, Siberian Branch Russian Academy of Sciences, 660036 Krasnoyarsk, Russia}
\affiliation{Digital Materials LLC, Kutuzovskaya str., 4A, 143001, Odintsovo, Russia}

\date{\today}

\begin{abstract}
Chiral magnetic textures, such as skyrmions, are of great interest to the condensed matter community due to their novel transport properties. The stabilization of topologically non-trivial magnetic phases, like the skyrmion lattice in cubic MnSi, is governed by underlying magnetic interactions which can be probed via measurements of spin-wave excitations. Here, we report high-resolution inelastic neutron scattering (INS) measurements of the spin waves in Fe-doped \mfs\ deep within its field-polarized ferromagnetic state. We observe non-reciprocal spin waves with a parabolic dispersion that shifts linearly with magnetic field. Crucially, the spin-wave stiffness is highly anisotropic, with values of 17.2(5) meV~\AA$^2$ parallel to the applied field and 9.3(3) meV~\AA$^2$ perpendicular to it. While the applied field fundamentally lowers the systems symmetry from cubic to uniaxial, we demonstrate that the induced spin-wave anisotropy is significantly large, which is not captured within the simple theoretical approximations widely applied up to date.

\end{abstract}
\maketitle

\section{Introduction}

Magnetic textures with non-trivial topology, such as skyrmions and chiral domain walls, have emerged as critical elements for the next-generation spintronic applications, which include devices for dense data storage and neuromorphic computing~\cite{back20202020, tomasello2014strategy}. To achieve precise control over the stability and dynamics of these textures for technological deployment, a foundational understanding of their underlying physics is essential. The key role is played by a delicate balance between the competing energy terms: the antisymmetric Dzyaloshinskii-Moriya interaction (DMI), which promotes chiral twisting; the symmetric exchange interaction, favoring uniform alignment; and magnetic anisotropy, which pins the magnetization to certain preferred axes~\cite{dhital2017exploring, maleyev2006cubic}. However, the precise quantification of these key interactions remains a significant challenge for many materials of interest.

A playground system for exploring this peculiar physics is the $B20$-type chiral material MnSi~\cite{georgii2019helical}. It hosts a well-characterized skyrmion lattice phase near the magnetic ordering temperature $T_{\text{c}}$ and serves as a benchmark for the chiral magnetism~\cite{nakajima2017skyrmion, adams2011long, sukhanov2019giant}. Very detailed experimental studies, which are frequently conducted with inelastic neutron scattering (INS), have revealed rich spin dynamics across the magnetic phase diagram. These include the observation of Stoner continua-type excitations associated with itinerant nature of the Mn moments~\cite{ishikawa1977magnetic}, emergent Landau levels within the skyrmion phase~\cite{weber2022topological} and the non-reciprocal magnons in the high-field polarized phase, where the excitation energy $\varepsilon(\mathbf{q}) \neq\ \varepsilon(\mathbf{-q})$~\cite{weber2018non, weber2019polarized}.

Chemical substitution provides powerful means to tune the underlying interactions. For instance, replacing Mn with Fe in the Mn$_{1-x}$Fe$_x$Si series systematically reduces the ordering temperature and contracts the pitch of the magnetic spiral~\cite{kindervater2020evolution, demishev2016quantum}. This behavior is attributed to a reduction of the ferromagnetic exchange constant $J$, as follows from the relation that defines the spiral propagation vector $|\mathbf{k_{\rm m}}| \propto\ D/J$. The ratio of $D/J$ thus sets the equilibrium period of chiral modulations~\cite{kugler2015band, glushkov2015scrutinizing, grigoriev2015spin}. Moreover, it was shown that $\mathbf{k}$ switches its direction from  $(\xi, \xi, \xi)$ to $(\xi, \xi, 0)$ at $x \approx\ 0.05$~\cite{bannenberg2018evolution, kindervater2020evolution}. The evolution of exchange interaction $J$ in the Mn$_{1-x}$Fe$_x$Si series was probed via systematic measurements of the spin-wave stiffness $A$, which is proportional to $J$, using small-angle neutron scattering (SANS) in applied magnetic fields~\cite{grigoriev2009helical}. Although allowing for a very good momentum resolution, this approach relies on certain approximations and provides only an indirect way to estimate the exchange interaction. In contrast, INS measurements access the spin-wave spectra directly, but can be hindered by insufficient momentum and energy resolution in the vicinity of $\mathbf{k}_{\rm m}$.

However, the resolution requirements can be eased if the system is driven into the fully-polarized state by strong enough external magnetic fields. This phase offers a classical well-defined starting point for analysis, as the spin-wave dispersion simplifies considerably. Namely, in this regime, the dispersion for a chiral magnet close to the minimum is usually described by:
\begin{align}
\epsilon_{\mathbf{q}} = A |\mathbf{q} - \mathbf{k}_{\mathrm{m}}|^2 + g\mu_{\rm B}(B-B_{\rm c}),
\label{Eq::dispersion}
\end{align}
where $g$ is the Land{\'e} $g$-factor, $\mu_{\rm B}$ is Bohr magneton, and $B_{\rm c}$ is the critical field needed to suppress the spin helix and induce the field-polarized state~\cite{grigoriev2015spin}. When this dispersion is resolved in an INS experiment, one can reliably extract $A$ as well $J$ via the relation $A = JSa^2$, where $S$ is the spin and $a$ is the lattice constant.

\begin{figure}[tb]
\center{\includegraphics[width=1\linewidth]{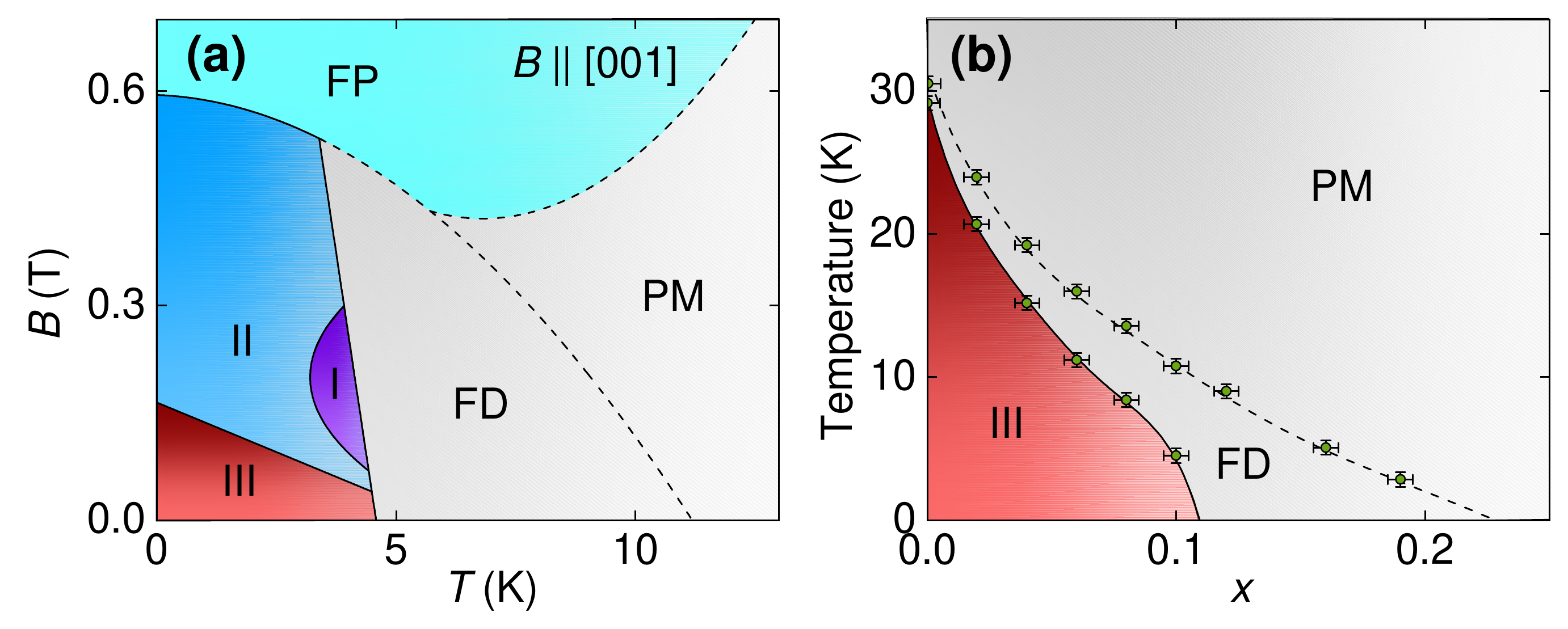}}
  \caption{~(a) The magnetic field--temperature phase diagram of \mfs.
  The phases I, II and III correspond to the incommensurate magnetic structures with the skyrmion-lattice, cone and helix order, respectively. FP is the field-polarized phase, PM is the paramagnetic phase, and FD is the fluctuation-disordered state.
  (b)~A compositional magnetic phase diagram for the solid solution Mn$_{1-x}$Fe$_x$Si for $B = 0$~T in coordinates of a temperature and the Fe content $x$. The dotted lines show the crossover from the FD to the PM phase.
  Data in both panels are reproduced from Ref.~\cite{kindervater2020evolution}. 
  }
  \label{fig_1}
  \vspace{-12pt}
\end{figure}

In the present work we apply such an approach and examine the spin excitations in \mfs. This compound represents an ideal balance for INS studies: it retains a sufficiently high magnetic ordering temperature $T_{\rm c} = 5.4$~K~\cite{bannenberg2018magnetization} and possesses the largest spiral propagation vector across the entire Mn$_{1-x}$Fe$_x$Si series ($\mathbf{k}_{\rm m}$$ = 0.07$~\AA$^{-1}$), which is crucial for achieving the momentum resolution needed to accurately resolve the the parabolic dispersion of Eq.~\eqref{Eq::dispersion}. Figures~\ref{fig_1}(a)--(b) illustrate the magnetic field-temperature phase diagram of \mfs\ and concentration-temperature phase diagram for Mn$_{1-x}$Fe$_x$Si series at zero-field magnetic field respectively~\cite{kindervater2020evolution}.

Our elastic neutron scattering data on \mfs\ confirm that in its ground state, the magnetic propagation vector is oriented along the $(\xi,\xi,0)$ direction, consistent with the prior studies~\cite{kindervater2020evolution}. We further demonstrate that a specific field protocol -- field-cooling under a high magnetic field, $\mathbf{B}\|[001]$, and subsequently reducing the field -- causes a reorientation of the spiral to a metastable state with $\mathbf{k}_{\rm m} \| (0,0,\xi)$, which is retained even upon returning to zero field. The INS data in the field-polarized phase reveal a well-defined parabolic dispersion with a gap that linearly scales with the field magnitude. The spin-wave stiffness $A$ along the applied-field direction shows only a negligible scaling with the field magnitude proving that the dispersion of Eq.~\eqref{Eq::dispersion} remains intact even for fields $B \gg B_{\rm c}$. While Eq.~\eqref{Eq::dispersion} implies an isotropic dispersion, we experimentally demonstrate that the spin-wave stiffness in \mfs\ perpendicular and parallel to the applied-field direction is essentially different. This makes the propagation of spin waves anisotropic, where the anisotropy is shaped by the applied field rather than the intrinsic crystalline symmetry. 

\section{Experiment}

A high-quality single crystal of \mfs\ was grown by the Czochralski method. The crystal quality was characterized by x-ray Laue diffraction and further in the present neutron measurements. The samples from the same batch were thoroughly characterized by bulk probes in earlier works~\cite{Stishov2024, Petrova2019}. The elastic and inelastic neutron scattering measurements were performed at the cold-neutron chopper spectrometer (CNCS)~\cite{CNCS1, CNCS2} at the SNS (ORNL Oak Ridge, USA) research facility. The crystal was mounted with its cubic [001] direction vertical, so that the $(hk0)$ plane coincided with the horizontal scattering plane of the instrument. Here $h$, $k$, and $l$ are the Miller indices labeling atomic planes in the crystal, and they define the wavevector transfer as $\mathbf{Q}=(h,k,l)$. A magnetic field was applied along the vertical direction  using an 8~T cryomagnet.
All measurements were conducted at base temperature of the cryomagnet, $T \approx\ 1.8$~K. The data were collected at the incident neutron energy of 3.32~meV, which yields the energy resolution of 0.11~meV at the elastic line.

\section{Results} \label{sec::results}

\subsection{Elastic neutron scattering}

\begin{figure}[tb]
\center{\includegraphics[width=1\linewidth]{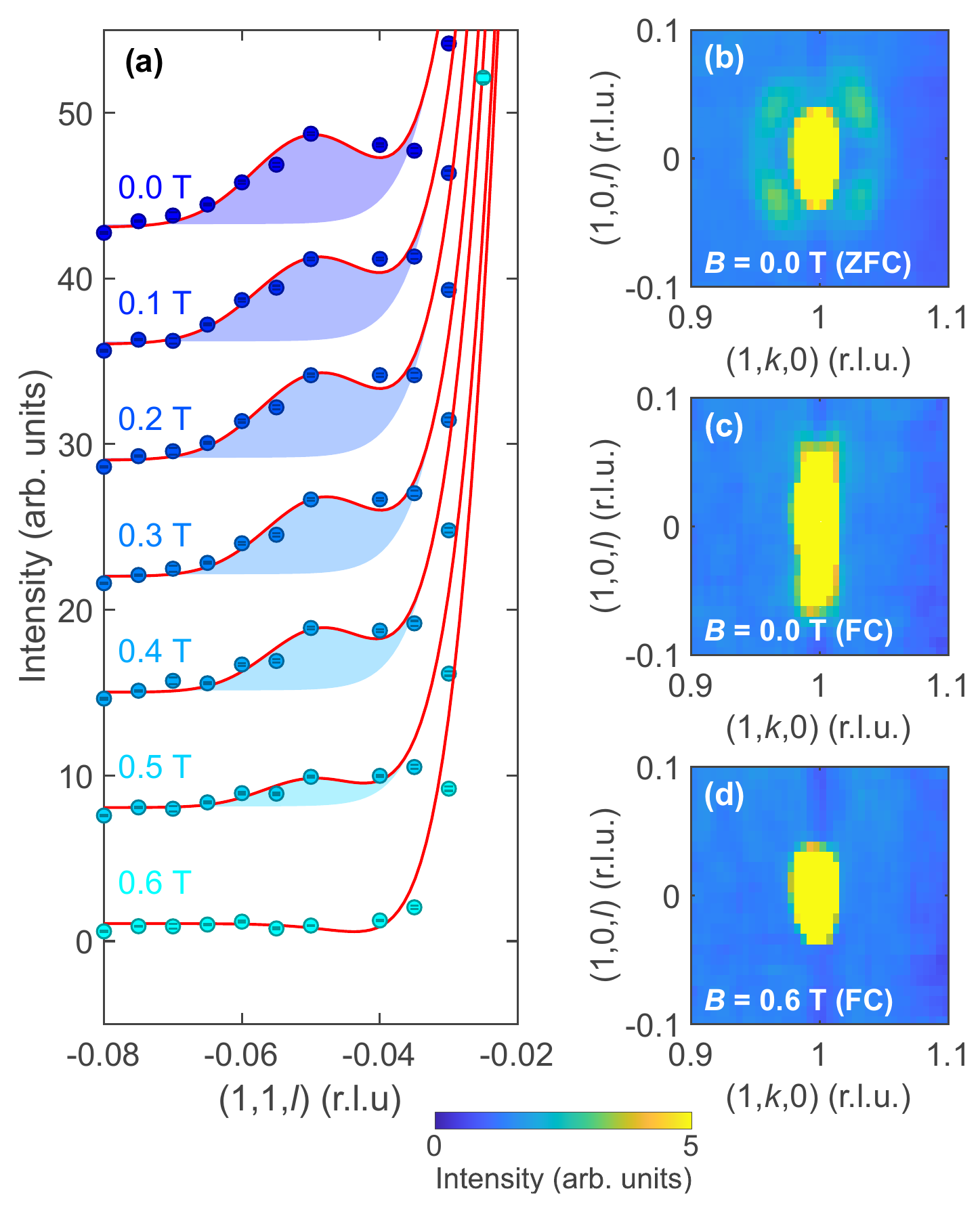}}
  \caption{~Elastic neutron scattering data. 
  (a)~Intensity profiles along the $(1,1,l)$ reciprocal direction collected after a field cooling. The dots show the experimental data, the red lines are fits to the data, the shaded areas represent the magnetic intensity (the large signal on the right side is due to the nuclear Bragg peak). Data are offset by 7 units for visual clarity.
  (b-d)~Elastic slices close to $(1,1,0)$ peak measured after different field protocols. 
  Data in panel (b) were collected after ZFC; in panel (d) after FC in 4~T and then reducing field to 0.6~T and in (c) after FC in 4~T field and reducing the field to 0~T after FC.
  All data were collected at base temperature of $T$=2 K.}
  \label{fig_2}
  \vspace{-12pt}
\end{figure}

We begin presentation of our results with the elastic neutron scattering data summarized in Figs.~\ref{fig_2}(a)--~\ref{fig_2}(d). Figure~\ref{fig_2}(b) shows a diffraction pattern collected at 2~K in the vicinity of the nuclear $(1,1,0)$ Bragg peak, after the sample is cooled in zero magnetic field. It shows four magnetic satellites which correspond to two arms of the spiral propagation vector $\mathbf{k}_{\rm m} = (\xi,\xi,0)$.
This differs from the behavior of pure MnSi, where the spiral propagates along the (111) direction~\cite{grigoriev2009helical, kindervater2020evolution, bannenberg2018evolution}. 

The application of a magnetic field gradually suppresses incommensurate order and only the nuclear peak can be seen in Fig.~\ref{fig_2}(d). When the magnetic field is decreased below 0.6~T, the magnetic reflections reappear again at $(1,1,\pm\xi)$ as shown in Fig.~\ref{fig_2}(c).  That means that the $\mathbf{k}_{\rm m}$ aligns parallel to the direction of the applied magnetic field (001), allowing for the formation of a single-domain magnetic state in the sample.

\begin{figure}[tb]
\center{\includegraphics[width=1\linewidth]{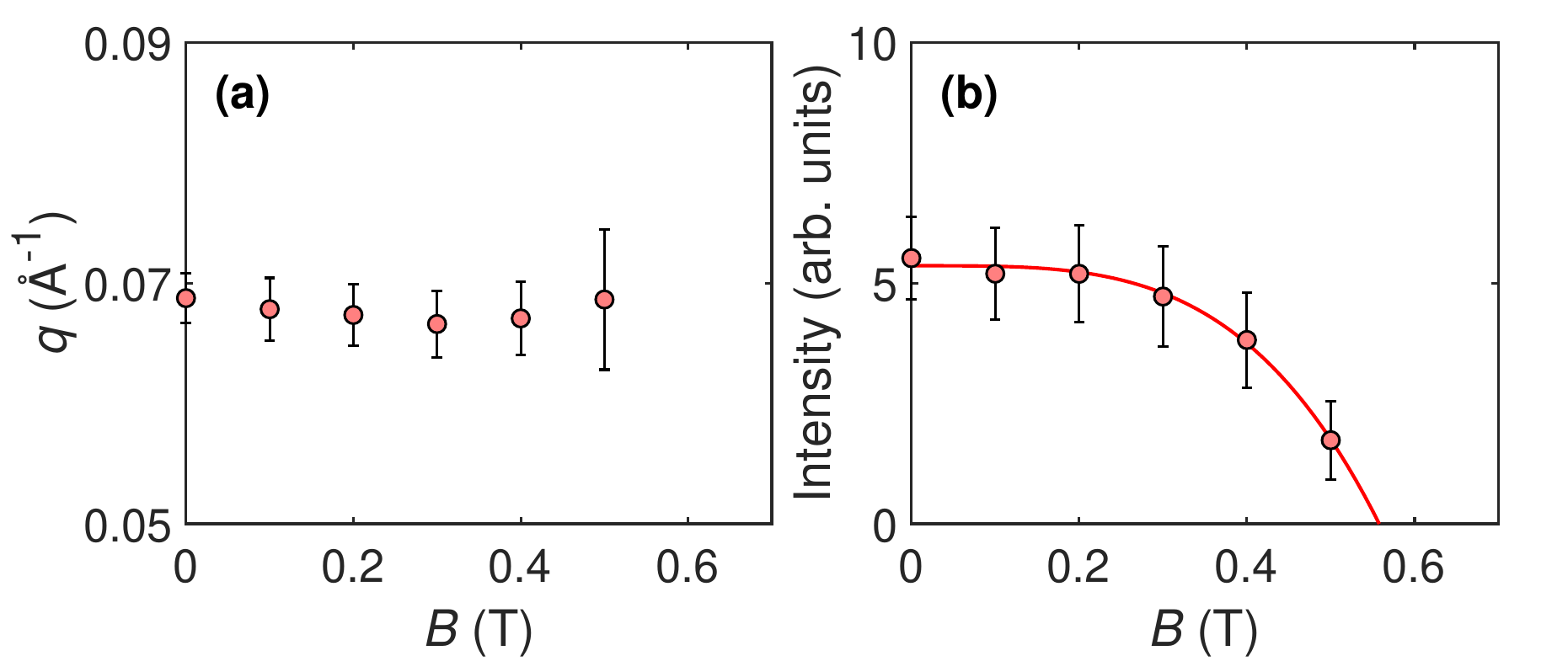}}
  \caption{~Magnetic-field dependence of the propagation vector (a) and the intensity of the magnetic reflection (b), measured at $T = 2$~K.
    These data were collected by first applying a magnetic field of $B = 4$~T, well above the critical field $B_{\rm c} = 0.6$~T, and then recording the data while reducing the field.
  }
  \label{fig_3}
  \vspace{-12pt}
\end{figure}

As the next step, we followed the field-induced evolution of the magnetic peak while decreasing the field from 0.6 to 0~T. The position and the intensity of the magnetic peaks were determined by fitting their profiles to a Gaussian function as shown in Fig.~\ref{fig_2}(a). The results of this analysis are summarized in Fig.~\ref{fig_3}. We found that the propagation vector does not change with the magnetic field and remains as $\mathbf{k}_{\rm m} = (0,0,0.050(1))$ (r.l.u.) or 0.069(2)~\AA$^{-1}$. The absence of any field-induced shift in $\mathbf{k}_{\rm m}$ demonstrates that the periodicity of the helimagnetic order is independent of the magnetic field up to its critical suppression value.

The magnetic field dependence of the elastic neutron scattering intensity, $I(B)$, was fitted by an empirical power-law function $I = I_0(1 - (B/B_{\rm c})^\alpha)$, where $B_{\rm c}$ is the critical field for the complete suppression of the incommensurate phase, $I_0$ is the intensity, and $\alpha$ is a fitting parameter. The results of the fitting procedure satisfactorily describe the experimental data and yield a critical field of $B_{\rm c}$ = 0.56(1)~T, where $\alpha \approx 3.6 $. The value of the critial field is in good agreement with the phase diagram [Fig.~\ref{fig_1}(a)] reported in Ref.~\cite{kindervater2020evolution}.

\section{Inelastic neutron scattering}

\begin{figure}[tb]
\center{\includegraphics[width=1\linewidth]{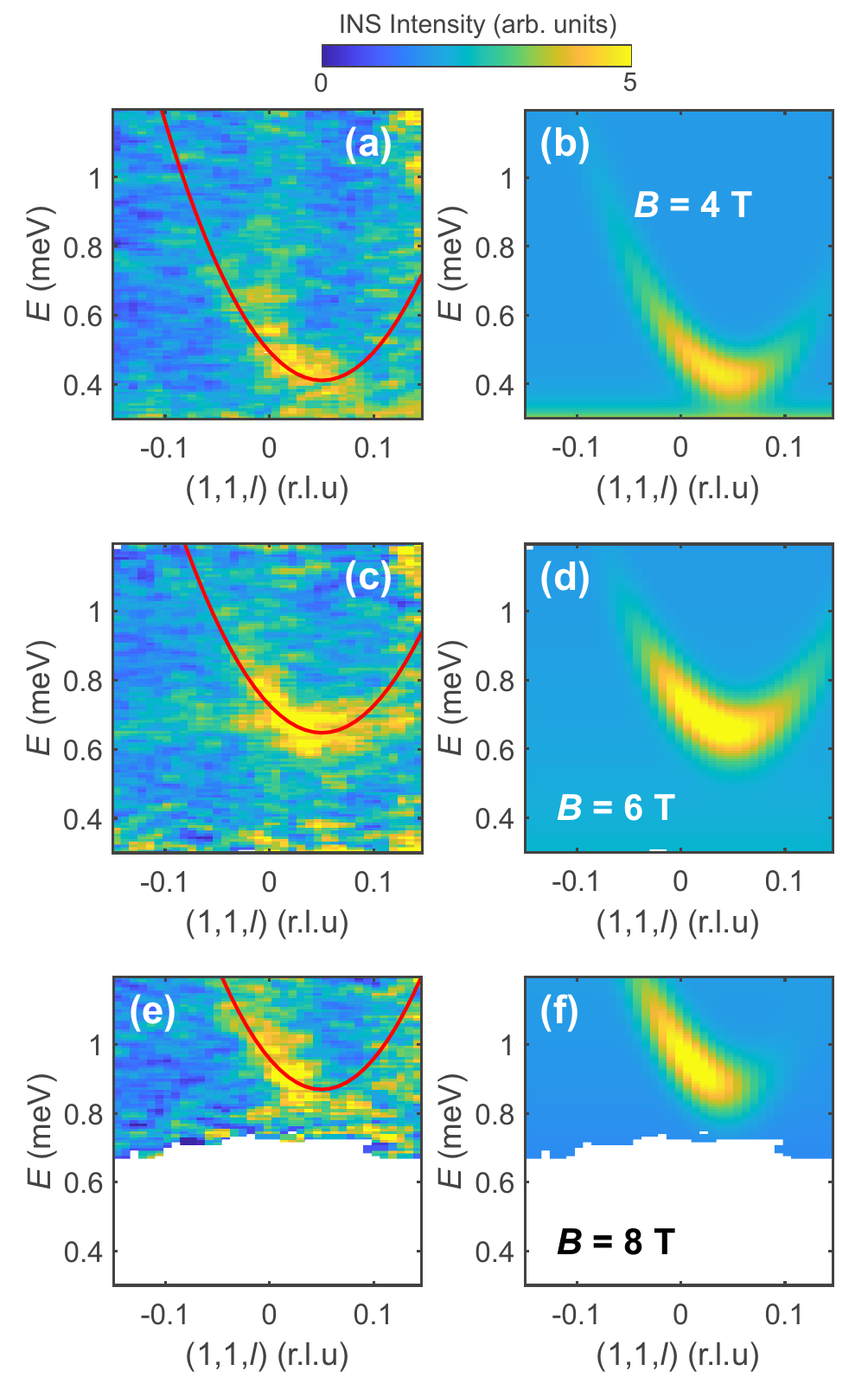}}
  \caption{~Inelastic neutron scattering spectra measured at $T = 2$~K in different magnetic fields. The experimental data are shown in panels (a,c,e) and the corresponding fits by Eq.~\eqref{Eq::2d_model} are represented in (b,d,f).
  All spectra are collected in vicinity of the $(1,1,0)$ nuclear Bragg peak along the $(11l)$ direction, which is parallel to the applied field.
  The red lines in (a,c,e) show the parabolic disperison of Eq.~\eqref{Eq::dispersion} calculated with the best fitted parameters (see main text and Fig.~\ref{fig_5}).
  }
  \label{fig_4}
  \vspace{-12pt}
\end{figure}

The inelastic neutron scattering spectra of Mn$_{1-x}$Fe$_x$Si measured at $T=2$~K under various magnetic fields are summarized in Fig.~\ref{fig_4}. The left panels [Fig.~\ref{fig_4}(a,c,e)] show the experimental data as color-coded intensity maps for momentum transfers parallel to the applied magnetic field, $\mathbf{B}\parallel(001)$. All spectra exhibit a well-defined magnon branch with a quadratic dispersion, consistent with the expected behavior of the field-polarized phase described by Eq.~\eqref{Eq::dispersion}.

The dispersion minimum is shifted from the $\Gamma$ point to the incommensurate wave vector $+\mathbf{k}_{\rm m}$, as clearly seen in Figs.~\ref{fig_4}(a,c,e). A similar shift has been observed in the field-polarized spectra of several systems with incommensurate magnetic ground states, including pristine MnSi~\cite{sato2016magnon} and GdRu$_2$Si$_2$~\cite{wood2025magnon}. Another striking feature is the absence of a corresponding magnon branch at $-\mathbf{k}_{\rm m}$. This nonreciprocal magnon behavior arises from the chiral Dzyaloshinskii--Moriya interaction in the spin Hamiltonian and has also been observed in parent MnSi~\cite{weber2019polarized} and $\alpha$-Cu$_2$V$_2$O$_7$~\cite{Gitgeatpong}.

The spectra exhibit sharp, coherent excitations near the dispersion minimum, while the spectral weight rapidly decreases and becomes barely visible for momenta $l>0.05$~r.l.u. We speculate that this pronounced loss of intensity may be associated with the proximity of the Stoner continuum~\cite{ishikawa1977magnetic,park2020momentum}.

For quantitative analysis of the spectra and extraction of the spin-wave stiffness and the energy gap, we developed an empirical model that approximates the scattering function $S(\mathbf{q}, \omega)$ as
\begin{align}
    S(\mathbf{q}, \omega) = a + b\omega + cI(\mathbf{q})\exp(-(\omega-\epsilon_{\mathbf{q}})^2/(2\sigma^2)),
    \label{Eq::2d_model}
\end{align}
where $\omega$ is the excitation frequency $(\omega=\frac{E}{\hbar})$, $b$ and $a$ are the slope and the constant term of the background, respectively, $c$ is the magnetic scale factor, $I(\mathbf{q})$ is the $\mathbf{q}$-dependent intensity profile, $\epsilon_{\mathbf{q}}$ is the dispersion relation, and $\sigma$ is the intrinsic peak width. Within this model, $I(\mathbf{q})$ was approximated by a rectangular function with Gaussian-smoothed edges, which yields a constant plateau between the two boundaries and the Gaussian tails, which describe the decay of intensity outside the plateau region. For each momenta $\mathbf{q}$, the corresponding spin-wave energy $\epsilon_{\mathbf{q}}$ was calculated according to Eq.~\eqref{Eq::dispersion}. 
Finally, the calculated intensity maps were convolved with a two-dimensional instrumental resolution function. The energy resolution was calculated using the Mantid software~\cite{Mantid}, while the momentum resolution along each direction was estimated from the width of the nuclear $(1,1,0)$ Bragg reflection. Because this reflection lies close to the reciprocal-space region investigated in the present measurements, it provides a suitable local estimate of the instrumental momentum resolution. Further details are provided in the Supplemental Material~\cite{SM}.
This empirical model accurately reproduces the experimental spectra and enables a reliable quantitative analysis of the spin-wave dispersion.


\begin{figure}[tb]
\center{\includegraphics[width=1\linewidth]{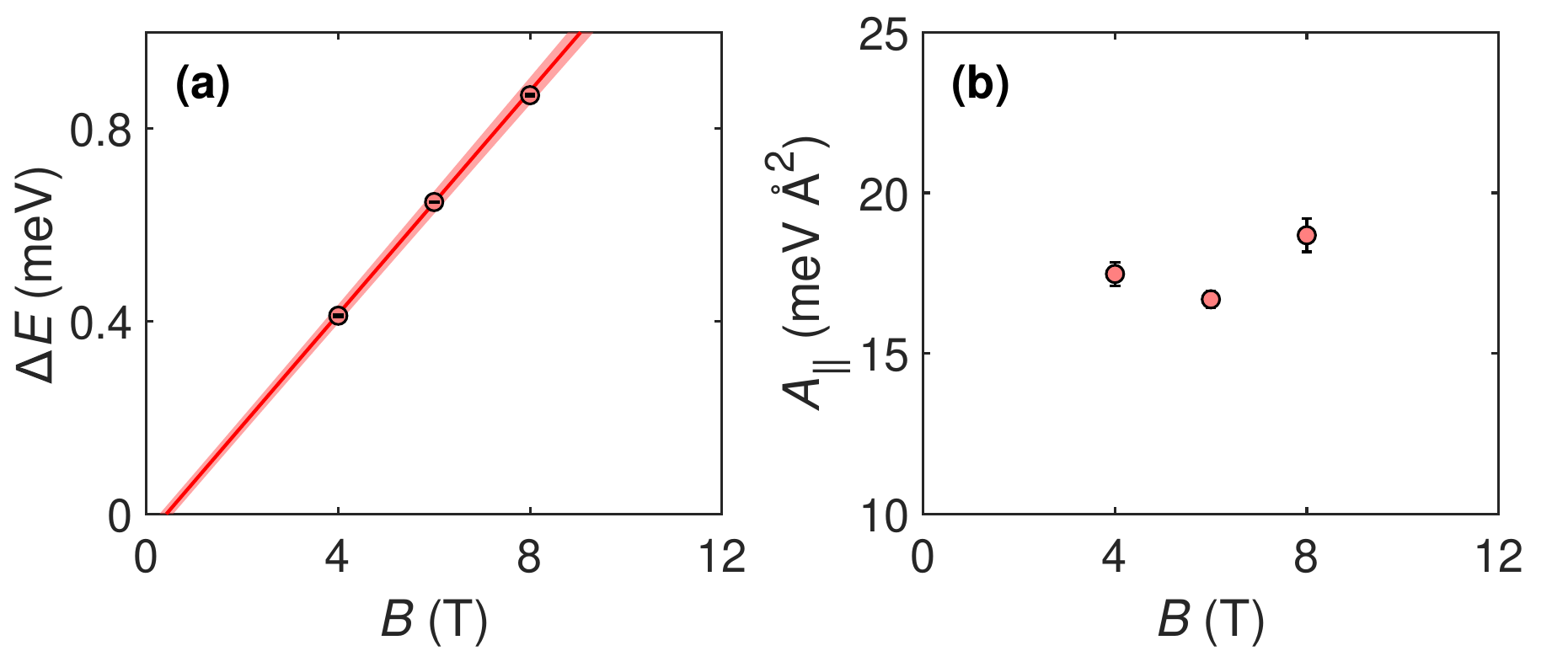}}
    \caption{~Magnetic-field dependence of (a) the spin-wave gap and (b) the longitudinal spin-wave stiffness, $A_{\parallel}$, measured at $T=2$~K. The red line in panel (a) shows the best linear fit to $\Delta E(B)$, while the light-shaded region represents the corresponding $1\sigma$ confidence band.
    }
  \label{fig_5}
  \vspace{-12pt}
\end{figure}

We fitted the model to the data collected at each applied magnetic field. For the datasets measured at 4 and 8~T, we fitted only the slices along the field direction, $(1,1,l)$. For the highest-quality dataset, measured at 6~T, we additionally analyzed slices along directions perpendicular to the field and performed a semi-global fit of the spin gap and spin-wave stiffness, as described below.

The fitted intensity maps for cuts along $(1,1,l)$ are shown in Figs.~\ref{fig_4}(b,d,f), demonstrating good agreement between the model and the experimental data. The fitted parameters are plotted as a function of the applied magnetic field $B$ in Fig.~\ref{fig_5}.
The spin-wave gap $\Delta E(B)$ increases linearly with magnetic field over the investigated field range. This behavior reflects the Zeeman-energy contribution, which is proportional to the field above the critical field $B_{\rm c}$~\cite{maleyev2006cubic,grigoriev2015spiral}. We fitted the experimental data using
\[
\Delta E(B)=g\mu_{\rm B}(B-B_{\rm c}).
\]
The fit yields $g=2.00(5)$ and $B_{\rm c}=0.41(14)$~T. The latter is close to $B_{\rm c}=0.56(1)$~T obtained from the elastic measurements [Fig.~\ref{fig_3}(b)] and to the value of approximately 0.6~T reported from thermodynamic measurements~\cite{kindervater2020evolution}.

The magnetic-field dependence of the longitudinal spin-wave stiffness $A_{\|}$, shown in Fig.~\ref{fig_5}(b), exhibits no significant variation and its mean value is $A_{\|}=17.2(5)$~meV~\AA$^2$.
The spin-wave stiffness is proportional to the exchange energy, $A\propto J$~\cite{grigoriev2009helical,grigoriev2015spin}, while increasing Fe substitution reduces the exchange energy~\cite{demishev2016quantum,glushkov2015scrutinizing}. The spin-wave stiffness obtained here is therefore expected to be smaller than that of pristine MnSi, for which $A=52(2)$~meV~\AA$^2$ at $T=5$~K~\cite{grigoriev2015spin,grigoriev2018spin}. Our result is consistent with this trend and with the previously reported values of $A=15.1$ and $18.7$~meV~\AA$^2$ at $T=3$~K for Mn$_{1-x}$Fe$_x$Si with $x=0.09$ and $0.11$, respectively~\cite{grigoriev2018spin}.

\begin{figure*}[!htb]
\centering
\includegraphics[width=\textwidth]{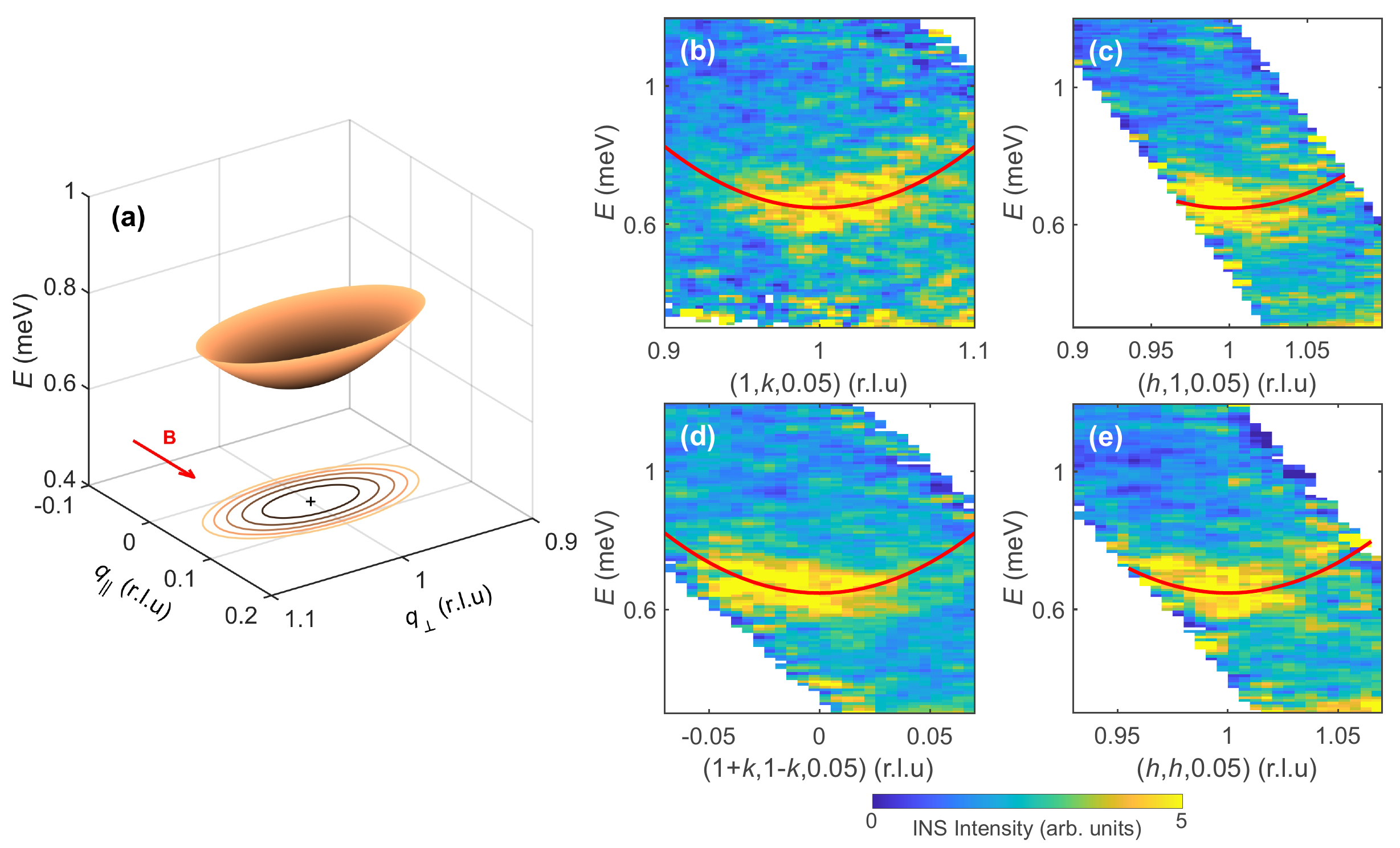}
\caption{~Spin wave dispersion at 6~T. 
(a) The 3D plot of the spin-wave dispersion in $(q_{\|}, q_{\perp})$ plane calculated using the fitted values. 
(b-e) The inelastic neutron scattering data along four reciprocal directions, which lie perpendicular to the applied field. 
The data were collected at $T = 2$~K.
The boundaries of the slices (d) and (e) were adjusted to ensure identical $\mathbf{Q}$-space spacings.
}
\label{fig_6}
\vspace{-12pt}
\end{figure*}

We also analyzed the inelastic neutron scattering spectra for wave vectors $\mathbf{q}$ perpendicular to the applied magnetic field using the 6~T dataset, which provides the highest signal-to-noise ratio [Figs.~\ref{fig_6}(b--e)].

The experimental data obtained for wave vectors $\mathbf{q}$ perpendicular to the applied magnetic field were fitted in two steps. First, all slices were fitted independently using the same spin-wave dispersion model as for the data with $\mathbf{q}\parallel\mathbf{B}$. In this analysis, the position of the dispersion minimum, $\mathbf{k}_{\rm m}$, was fixed at zero reduced momentum [corresponding to the absolute momentum $(1~1~0.05)$~r.l.u.], while the energy gap $E$ was fixed to the value obtained from the analysis of the $\mathbf{q}\parallel\mathbf{B}$ data [Fig.~\ref{fig_4}(c,d)].
The resulting spin-wave stiffnesses are 8.3(6), 7(1), 8.5(5), and 6.4(9)~meV~\AA$^2$ for the slices shown in Figs.~\ref{fig_6}(b--e), respectively. Although these values are not identical, they are mutually consistent within their uncertainties. Their mean value, $A_{\perp}=7.6(4)$~meV~\AA$^2$, is substantially smaller than the longitudinal stiffness, $A_{\|}=14.7(4)$~meV~\AA$^2$, which was also obtained from an independent fit.

Having established that the four independently determined values of $A_{\perp}$ are mutually consistent, we performed a semi-global fit of all five slices measured at 6~T. In this analysis, the energy gap $E$ and the transverse spin-wave stiffness $A_{\perp}$ were treated as global parameters, while the remaining parameters were allowed to vary independently for each slice. The fit yields $A_{\perp}=9.3(3)$, $A_{\|}=16.7(3)$~meV~\AA$^2$, and $\Delta E=0.6476(9)$~meV, providing an accurate and robust empirical description of the experimental data.
The pronounced difference between $A_{\|}$ and $A_{\perp}$ demonstrates a substantial field-induced anisotropy of the spin-wave propagation.

Using the dispersion parameters obtained for momentum transfers parallel and perpendicular to the applied magnetic field, we reconstructed the three-dimensional dispersion surface $E(q_{\perp},q_{\|})$, as shown in Fig.~\ref{fig_6}(a). This surface provides a direct visualization of the pronounced anisotropy of the spin-wave dispersion.

\section{Discussion and Conclusion}

We begin the discussion of the observed anisotropy of the spin-wave dispersion in the field-polarized state with a brief description of the magnon anisotropy previously found in the incommensurate phase of MnSi. 
In the low-energy limit, the magnetic system can be described by a free energy density, which includes several contributions: exchange interaction, DMI, Zeeman energy in an external magnetic field, as well as the dipole interaction between magnetic moments~\cite{garst2017collective, weber2018field}. This model was successfully used to describe the experimentally observed magnon dispersion at zero and low-field states~\cite{weber2018field, kugler2015band}.

The helical magnetic order in MnSi can be viewed as a one-dimensional magnetic crystal with a period given by the spiral length $\propto1/|\mathbf{k}_{\rm m}|$~\cite{garst2017collective}. The propagation of magnons in this phase is determined by the direction of their motion relative to the propagation wavevector. When magnons propagate strictly along the spiral axis, the periodic magnetic structure does not affect them - they move freely, and their energy increases smoothly (quadratically) with increasing wave vector. However, if the magnon momentum has a component perpendicular to the spiral axis, the propagating magnon begins to experience Bragg scattering off the periodic structure~\cite{garst2017collective}. For sufficiently large values of the perpendicular component, the low-energy bands lose their dependence on the momentum along the spiral axis and become flat. Thus, in the helical phase, the magnon spectrum exhibits pronounced anisotropy: dispersion along and perpendicular to the spiral axis is qualitatively different.

This theoretical approach for describing noncentrosymmetric cubic magnets in the helical phase was confirmed in the experimental work~\cite{kugler2015band} using inelastic neutron scattering on a monodomain crystal of MnSi. By employing different scattering geometries, the authors measured the magnon dispersion for the wave-vectors directed along the spiral axis but with different transverse components.   
An analysis of the obtained data revealed the formation of the flat bands and a pronounced anisotropy of the magnon dispersion, which is in full agreement with the theoretical predictions.


At magnetic fields above the critical value, $B>B_{\rm c}$, the helical magnetic order is suppressed, and a collinear field-polarized (FP) state is established in the system. Within the theoretical framework for describing cubic magnets, as mentioned above, the FP state is described by Eq.~\eqref{Eq::dispersion}. The dispersion is predicted to be isotropic meaning that the spin-wave stiffness should not depend on whether the magnon propagates along the field, or perpendicular to it. Such an approximation assumes that the applied field induces the spin anisotropy only via the Zeeman term. Fundamentally, the field applied along the [001] crystal axis reduces the cubic symmetry to uniaxial, which allows for higher-order terms in the dispersion relation. While this should inevitably induce a finite spin-wave stiffness anisotropy beyond the simple form of Eq.~\eqref{Eq::dispersion}, its significance and the order of magnitude in a real material remained unclear.

Here, our experimental data reveals a clear anisotropic spin-wave behavior. As shown in Fig.~\ref{fig_6}(a), the spin-wave stiffness values extracted from the magnon dispersion differ significantly for magnon propagation directions along and perpendicular to the magnetic field. The stiffness ratio is $A_{\|}/A_{\perp} \approx\ 2$, indicating a strong field-induced anisotropy. Such a pronounced anisotropic response to the symmetry-lowering applied field is not accounted for in the standard theory of the polarized phase of a cubic chiral magnet~\cite{maleyev2006cubic, kataoka1987spin}, which has so far been deemed fully applicable in the real materials~\cite{grigoriev2015spin,grigoriev2015spiral,grigoriev2018spin}.


We consider few possible ingredients that may be relevant to the observed dispersion. 
The first one is the single-ion anisotropy. Cubic anisotropy in the polarized phase was considered in Ref.~\cite{maleyev2006cubic}, it was shown to primarily affects the energy gap and does not alter the spin-wave stiffness, thus it cannot induce the observed effect. Moreover, the observed spin-wave anisotropy involves the crystal axes that are indistinguishable for the cubic anisotropy, i.e. [100], [010], and [001].

Another plausible contribution is the anisotropic exchange. However, it was shown that it might lead to a minor variation of the propagation vector in the helical phase for the symmetry inequivalent axes, such as [100] and [110] in the cubic helimagnets, akin to the single-ion anisotropy~\cite{Ukleev_1,Ukleev_2}. Therefore, it is also not expected to cause the pronounced anisotropy of the spin-wave stiffness, as observed here.


The dipole-dipole interaction is known to cause some spin-wave anisotropy with respect to the applied field~\cite{maleyev2006cubic}; however, having in mind a very low magnetic moment in \mfs, its magnitude is unlikely to explain the observed ratio of $A_{\|}/A_{\perp} \approx\ 2$ either. Especially in the bulk sample, as studied here, in which the stray fields are much smaller than in thin-film samples of ferromagnets, where a noticeable dipolar contributions is typically observed~\cite{dipolar_1, dipolar_2}. 
We support these statements with a numerical estimate of the effect of dipolar coupling and exchange anisotropy on the spin wave anisotropy in SM~\cite{SM}.


The role of the inherent structural disorder in Mn$_{1-x}$Fe$_x$Si, which modifies the phase diagram and possibly causes a formation of the quantum Griffiths phase in certain regions of the $B-T$ phase diagram~\cite{demishev2016magnetization, mishra2023quantum}, remains an open question. While the relation between the disorder and the spin-wave stiffness was not yet thoroughly addressed in the theory studies, it remains unlikely that the isotropic disorder can cause a spin-wave anisotropy. The main known mechanism behind impurities is the renormalization of the propagation vector~\cite{Utesov_1}, which is indeed present in \mfs.

Since \mfs~is a highly itinerant magnet, its exchange interactions are driven by the properties of its Fermi surface. As was discussed in Ref.~\cite{FermiSurface}, the magnetic field applied to MnSi lowers the symmetry of the Fermi surfaces, making the electronic bands dispersing along the applied field, and perpendicular to it different. Such a symmetry reduction was shown to change the topological properties of the electronic structure, inducing the nodal planes at the Fermi surface that can be manipulated by the field direction. We speculate that the strong spin-wave anisotropy might also be driven by the symmetry reduction of the Fermi surface, although this scenario would require further verifications and should be considered hypothetical in light of our findings.

To conclude, we report elastic and inelastic neutron scattering study of \mfs. Our elastic neutron scattering data show a formation of the multi-domain helical order with the propagation wavevector along the (110) direction . This order can be suppressed in favor of the field-polarized state at $B_{\rm c} = 0.56(1)$~T (for $\mathbf{B} \| [001]$). We show that after field cycling above $B_{\rm c}$ and returning to zero field the magnetic helix reorients with $\mathbf{k}_{\rm m} \| [001]$ and remains in this single-domain metastable state. Our inelastic data revealed the spin-wave mode with the quadratic dispersion, which shifts linearly with the applied field due to Zeeman energy gap. The spectra show unexpected anisotropy of the spin-wave stiffness for the wavevectors along and perpendicular to the applied magnetic field. Such an effect is not predicted by the standard theories of helical magnets and its appropriate theoretical description should be addressed in future studies.


\section{Acknowledgments}
We acknowledge stimulating discussions S. V. Grigoriev and K. A. Pschenichnyi and thank N. D. Andriushin for the help with Laue measurements.
This research used resources at the Spallation Neutron Source, a DOE Office of Science User Facility operated by Oak Ridge National Laboratory. The beam time was allocated to CNCS on proposal number IPTS-20409.

\bibliography{bibliography}

@article{Stishov2024,
	author = {Stishov, S.M. and Petrova, A.E. and Menzel, Dirk and Belemuk, A.M.},
	title = {{Quantum criticality features in the Co, Fe doped MnSi}},
	year = {2024},
	journal = {Phys. B: Cond. Mat.},
	volume = {675},
    pages ={415607},
	doi = {10.1016/j.physb.2023.415607},
	type = {Article},
	source = {Scopus}
}

@article{Petrova2019,
	author = {Petrova, A.E. and Gavrilkin, S. Yu. and Menzel, Dirk and Stishov, S.M.},
	title = {{Physical properties of (Mn$_{1-x}$Fe$_x$)Si at $x \simeq0.15$ along the critical trajectory}},
	year = {2019},
	journal = {Phys. Rev. B},
	volume = {100},
	number = {9},
    pages = {094403},
	doi = {10.1103/PhysRevB.100.094403},
	source = {Scopus},
}

@article{Ukleev_1,
  title = {{Competing anisotropies in the chiral cubic magnet Co$_8$Zn$_8$Mn$_4$ unveiled by resonant x-ray magnetic scattering}},
  author = {Ukleev, Victor and Utesov, Oleg I. and Luo, Chen and Radu, Florin and Wintz, Sebastian and Weigand, Markus and Finizio, Simone and Winter, Moritz and Tahn, Alexander and Rellinghaus, Bernd and Karube, Kosuke and Tokura, Yoshinori and Taguchi, Yasujiro and White, Jonathan S.},
  journal = {Phys. Rev. B},
  volume = {109},
  issue = {18},
  pages = {184415},
  numpages = {7},
  year = {2024},
  month = {May},
  publisher = {American Physical Society},
  doi = {10.1103/PhysRevB.109.184415},
  url = {https://link.aps.org/doi/10.1103/PhysRevB.109.184415}
}

@article{Ukleev_2,
  title = {Signature of anisotropic exchange interaction revealed by vector-field control of the helical order in a FeGe thin plate},
  author = {Ukleev, Victor and Utesov, Oleg and Yu, Le and Luo, Chen and Chen, Kai and Radu, Florin and Yamasaki, Yuichi and Kanazawa, Naoya and Tokura, Yoshinori and Arima, Taka-hisa and White, Jonathan S.},
  journal = {Phys. Rev. Res.},
  volume = {3},
  issue = {1},
  pages = {013094},
  numpages = {8},
  year = {2021},
  month = {Jan},
  publisher = {American Physical Society},
  doi = {10.1103/PhysRevResearch.3.013094},
  url = {https://link.aps.org/doi/10.1103/PhysRevResearch.3.013094}
}

@article{Utesov_1,
  title = {Cubic B20 helimagnets with quenched disorder in magnetic field},
  author = {Utesov, O. I. and Syromyatnikov, A. V.},
  journal = {Phys. Rev. B},
  volume = {99},
  issue = {13},
  pages = {134412},
  numpages = {8},
  year = {2019},
  month = {Apr},
  publisher = {American Physical Society},
  doi = {10.1103/PhysRevB.99.134412},
  url = {https://link.aps.org/doi/10.1103/PhysRevB.99.134412}
}

@article{FermiSurface,
	author = {Wilde, Marc A. and Dodenh{\"o}ft, Matthias and Niedermayr, Arthur and Bauer, Andreas and Hirschmann, Moritz M. and Alpin, Kirill and Schnyder, Andreas P. and Pfleiderer, Christian},
	date = {2021/06/01},
	doi = {10.1038/s41586-021-03543-x},
	id = {Wilde2021},
	isbn = {1476-4687},
	journal = {Nature},
	number = {7863},
	pages = {374--379},
	title = {Symmetry-enforced topological nodal planes at the Fermi surface of a chiral magnet},
	url = {https://doi.org/10.1038/s41586-021-03543-x},
	volume = {594},
	year = {2021}}

@article{dipolar_1,
  title = {Spin-wave focusing induced by dipole-dipole interaction in synthetic antiferromagnets},
  author = {Gallardo, R. A. and Alvarado-Seguel, P. and K\'akay, A. and Lindner, J. and Landeros, P.},
  journal = {Phys. Rev. B},
  volume = {104},
  issue = {17},
  pages = {174417},
  numpages = {12},
  year = {2021},
  month = {Nov},
  publisher = {American Physical Society},
  doi = {10.1103/PhysRevB.104.174417},
  url = {https://link.aps.org/doi/10.1103/PhysRevB.104.174417}
}

@article{dipolar_2,
  title = {Reconfigurable Spin-Wave Nonreciprocity Induced by Dipolar Interaction in a Coupled Ferromagnetic Bilayer},
  author = {Gallardo, R.A. and Schneider, T. and Chaurasiya, A.K. and Oelschl\"agel, A. and Arekapudi, S.S.P.K. and Rold\'an-Molina, A. and H\"ubner, R. and Lenz, K. and Barman, A. and Fassbender, J. and Lindner, J. and Hellwig, O. and Landeros, P.},
  journal = {Phys. Rev. Appl.},
  volume = {12},
  issue = {3},
  pages = {034012},
  numpages = {11},
  year = {2019},
  month = {Sep},
  publisher = {American Physical Society},
  doi = {10.1103/PhysRevApplied.12.034012},
  url = {https://link.aps.org/doi/10.1103/PhysRevApplied.12.034012}
}

@article{weber2019polarized,
  title={{Polarized inelastic neutron scattering of nonreciprocal spin waves in MnSi}},
  author={T. Weber and J. Waizner and P. Steffens and A. Bauer and C. Pfleiderer and M. Garst and P. B{\"o}ni},
  journal={Phys. Rev. B},
  volume={100},
  number={6},
  pages={060404},
  year={2019},
  publisher={APS},
  doi={10.1103/PhysRevB.100.060404}
}

@article{grigoriev2015spin,
  title={{Spin waves in full-polarized state of Dzyaloshinskii-Moriya helimagnets: Small-angle neutron scattering study}},
  author={S. V. Grigoriev and A. S. Sukhanov and E. V. Altynbaev and S. A. Siegfried and A. Heinemann and P. Kizhe and S. V. Maleyev},
  journal={Phys. Rev. B},
  volume={92},
  number={22},
  pages={220415},
  year={2015},
  publisher={APS},
  doi={10.1103/PhysRevB.92.220415}
}

@article{kugler2015band,
  title={{Band structure of helimagnons in MnSi resolved by inelastic neutron scattering}},
  author={M. Kugler and G. Brandl and J. Waizner and M. Janoschek and R. Georgii and A. Bauer and K. Seemann and A. Rosch and C. Pfleiderer and P. B{\"o}ni and M. Garst},
  journal={Phys. Rev. Lett.},
  volume={115},
  number={9},
  pages={097203},
  year={2015},
  publisher={APS},
  doi={10.1103/PhysRevLett.115.097203}
}

@article{weber2022topological,
  title={Topological magnon band structure of emergent Landau levels in a skyrmion lattice},
  author={T. Weber and D. M. Fobes and J. Waizner and P. Steffens and G. S. Tucker and M. B{\"o}hm and L. Beddrich and C. Franz and H. Gabold and R. Bewley and D. Voneshen and M. Skoulatos and R. Georgii and G. Ehlers and A. Bauer and C. Pfleiderer and P. B{\"o}ni and M. Janoschek  and M. Garst},
  journal={Science},
  volume={375},
  number={6584},
  pages={1025--1030},
  year={2022},
  publisher={American Association for the Advancement of Science},
  doi={10.1126/science.abe4441}
}

@article{weber2018field,
  title={Field dependence of nonreciprocal magnons in chiral MnSi},
  author={T. Weber and J. Waizner and G. S. Tucker and R. Georgii and M. Kugler and A. Bauer and C. Pfleiderer and M. Garst and P. B{\"o}ni},
  journal={Phys. Rev. B},
  volume={97},
  number={22},
  pages={224403},
  year={2018},
  publisher={APS},
  doi={10.1103/PhysRevB.97.224403}
}

@article{bannenberg2018evolution,
  title={{Evolution of helimagnetic correlations in Mn$_{1-x}$Fe$_x$Si with doping: A small-angle neutron scattering study}},
  author={L. J. Bannenberg and R. M. Dalgliesh and T. Wolf and F. Weber and C. Pappas},
  journal={Phys. Rev. B},
  volume={98},
  number={18},
  pages={184431},
  year={2018},
  publisher={APS},
  doi={10.1103/PhysRevB.98.184431}       
}

@article{kindervater2020evolution,
  title={{Evolution of magnetocrystalline anisotropies in Mn$_{1-x}$Fe$_x$Si and M$_{1-x}$Co$_x$Si as inferred from small-angle neutron scattering and bulk properties}},
  author={J. Kindervater and T. Adams and A. Bauer and F. X. Haslbeck and A. Chacon and S. M{\"u}hlbauer and F. Jonietz and A. Neubauer and U. Gasser and G. Nagy and N. Martin and W. H{\"a}u{\ss}ler and R. Georgii and M. Garst and and C. Pfleiderer},
  journal={Phys. Rev. B},
  volume={101},
  number={10},
  pages={104406},
  year={2020},
  publisher={APS},
  doi={10.1103/PhysRevB.101.104406}
}

@article{grigoriev2009helical,
  title={{Helical spin structure of Mn$_{1-y}$Fe$_y$Si under a magnetic field: Small angle neutron diffraction study}},
  author={Grigoriev, S. V. and Dyadkin, V. A. and Moskvin, E. V. and Lamago, D and Wolf, Th and Eckerlebe, H and Maleyev, S. V.},
  journal={Phys. Rev. B},
  volume={79},
  number={14},
  pages={144417},
  year={2009},
  publisher={APS},
  doi={10.1103/PhysRevB.79.144417}
}

@article{sato2016magnon,
  title={{Magnon dispersion shift in the induced ferromagnetic phase of noncentrosymmetric MnSi}},
  author={Taku J. Sato and D. Okuyama and T. Hong and A. Kikkawa and Y. Taguchi and T. Arima and Y. Tokura},
  journal={Phys. Rev. B},
  volume={94},
  number={14},
  pages={144420},
  year={2016},
  publisher={APS},
  doi={10.1103/PhysRevB.94.144420}
}

@article{dhital2017exploring,
  title={{Exploring the origins of the Dzyaloshinskii-Moriya interaction in MnSi}},
  author={C. Dhital and L. DeBeer-Schmitt and Q. Zhang and W. Xie and D. P. Young and J. F. DiTusa},
  journal={Phys. Rev. B},
  volume={96},
  number={21},
  pages={214425},
  year={2017},
  publisher={APS},
  doi={10.1103/PhysRevB.96.214425}
}

@article{maleyev2006cubic,
  title={{Cubic magnets with Dzyaloshinskii-Moriya interaction at low temperature}},
  author={S. V. Maleyev},
  journal={Phys. Rev. B},
  volume={73},
  number={17},
  pages={174402},
  year={2006},
  publisher={APS},
  doi={10.1103/PhysRevB.73.174402}
}

@article{georgii2019helical,
  title={{The Helical Magnet MnSi: Skyrmions and Magnons}},
  author={R. Georgii and T. Weber},
  journal={Quantum Beam Sci.},
  volume={3},
  number={1},
  pages={4},
  year={2019},
  publisher={MDPI},
  doi={10.3390/qubs3010004}
}

@article{nakajima2017skyrmion,
  title={{Skyrmion lattice structural transition in MnSi}},
  author={T. Nakajima and H. Oike and A. Kikkawa and E. P. Gilbert and N. Booth and K. Kakurai and Y. Taguchi and Y. Tokura and F. Kagawa and T. Arima},
  journal={Sci. Adv.},
  volume={3},
  number={6},
  pages={e1602562},
  year={2017},
  publisher={American Association for the Advancement of Science},
  doi={10.1126/sciadv.1602562}
}

@article{adams2011long,
  title={{Long-range crystalline nature of the skyrmion lattice in MnSi}},
  author={T. Adams and S. M{\"u}hlbauer and C. Pfleiderer and F. Jonietz and A. Bauer and A. Neubauer and R. Georgii and P. B{\"o}ni and U. Keiderling and K. Everschor and others},
  journal={Physical review letters},
  volume={107},
  number={21},
  pages={217206},
  year={2011},
  publisher={APS},
  doi={10.1103/PhysRevLett.107.217206}
}

@article{ishikawa1977magnetic,
  title={{Magnetic excitations in the weak itinerant ferromagnet MnSi}},
  author={Y. Ishikawa and G. Shirane and J. A. Tarvin and M. Kohgi},
  journal={Phys. Rev. B},
  volume={16},
  number={11},
  pages={4956},
  year={1977},
  publisher={APS},
  doi={10.1103/PhysRevB.16.4956}
}

@article{demishev2016quantum,
  title={Quantum phase transitions in spiral magnets without an inversion center},
  author={S. V. Demishev and V. V. Glushkov and S. V. Grigoriev and M. I. Gilmanov and I. I. Lobanova and A. N. Samarin and A. V. Semeno and N. E. Sluchanko},
  journal={Phys. Usp.},
  volume={59},
  number={6},
  pages={559},
  year={2016},
  publisher={IOP Publishing},
  doi={10.3367/UFNe.2016.02.037767}
}

@article{glushkov2015scrutinizing,
  title={{Scrutinizing Hall Effect in Mn$_{1-x}$Fe$_x$Si: Fermi Surface Evolution and Hidden Quantum Criticality}},
  author={Glushkov, V. V. and Lobanova, I. I. and Ivanov, V. Yu. and Voronov, V. V. and Dyadkin, V. A. and Chubova, N. M. and Grigoriev, S. V. and Demishev, S. V.},
  journal={Phys. Rev. Lett.},
  volume={115},
  number={25},
  pages={256601},
  year={2015},
  publisher={APS},
  doi={10.1103/PhysRevLett.115.256601}
}

@article{weber2018non,
  title={{Non-reciprocal magnons in non-centrosymmetric MnSi}},
  author={Weber, Tobias and Waizner, Johannes and Tucker, G. S. and Beddrich, Lukas and Skoulatos, Markos and Georgii, Robert and Bauer, Andreas and Pfleiderer, Christian and Garst, Markus and B{\"o}ni, Peter},
  journal={AIP Adv.},
  volume={8},
  number={10},
  year={2018},
  publisher={AIP Publishing},
  doi={10.1063/1.5041036}
}

@article{back20202020,
  title={The 2020 skyrmionics roadmap},
  author={Back, Christian and Cros, Vincent and Ebert, Hubert and Everschor-Sitte, Karin and Fert, Albert and Garst, Markus and Ma, Tianping and Mankovsky, Sergiy and Monchesky, T. L. and Mostovoy, Maxim and others},
  journal={J. Phys. D: Appl. Phys.},
  volume={53},
  number={36},
  pages={363001},
  year={2020},
  publisher={IOP Publishing},
  doi={10.1088/1361-6463/ab8418}
}

@article{tomasello2014strategy,
  title={A strategy for the design of skyrmion racetrack memories},
  author={Tomasello, Riccardo and Martinez, E and Zivieri, Roberto and Torres, Luis and Carpentieri, Mario and Finocchio, Giovanni},
  journal={Sci. Rep.},
  volume={4},
  number={1},
  pages={6784},
  year={2014},
  publisher={Nature Publishing Group UK London},
  doi={10.1038/srep06784}
}

@article{grigoriev2015spiral,
  title={{From spiral to ferromagnetic structure in B20 compounds: Role of cubic anisotropy}},
  author={Grigoriev, S. V. and Sukhanov, A. S. and Maleyev, S. V.},
  journal={Phys. Rev. B},
  volume={91},
  number={22},
  pages={224429},
  year={2015},
  publisher={APS},
  doi={10.1103/PhysRevB.91.224429}
}

@article{wood2025magnon,
  title={{A magnon band analysis of GdRu$_2$Si$_2$ in the field-polarized state}},
  author={Wood, G. D. A. and Stewart, J. R. and Mayoh, D. A. and Paddison, J. A. M. and Bouaziz, J and Tobin, S. M and Petrenko, O. A and Lees, M. R and Manuel, P and Staunton, J. B. and others},
  journal={npj Quantum Mater.},
  volume={10},
  number={1},
  pages={39},
  year={2025},
  publisher={Nature Publishing Group UK London},
  doi={10.1038/s41535-025-00755-6}
}

@article{Gitgeatpong,
  title = {{Nonreciprocal Magnons and Symmetry-Breaking in the Noncentrosymmetric Antiferromagnet}},
  author = {Gitgeatpong, G. and Zhao, Y. and Piyawongwatthana, P. and Qiu, Y. and Harriger, L. W. and Butch, N. P. and Sato, T. J. and Matan, K.},
  journal = {Phys. Rev. Lett.},
  volume = {119},
  pages = {047201},
  numpages = {5},
  year = {2017},
  doi = {10.1103/PhysRevLett.119.047201}
}

@article{grigoriev2018spin,
  title={{Spin-wave stiffness in the Dzyaloshinskii-Moriya helimagnets Mn$_{1-x}$Fe$_x$Si}},
  author={Grigoriev, S. V. and Altynbaev, E. V. and Siegfried, S-A and Pschenichnyi, K. A. and Menzel, D and Heinemann, A and Chaboussant, G},
  journal={Phys. Rev. B},
  volume={97},
  number={2},
  pages={024409},
  year={2018},
  publisher={APS},
  doi={10.1103/PhysRevB.97.024409}
}

@article{kataoka1987spin,
  title={Spin waves in systems with long period helical spin density waves due to the antisymmetric and symmetric exchange interactions},
  author={Kataoka, Mitsuo},
  journal={J. Phys. Soc. Jpn.},
  volume={56},
  number={10},
  pages={3635--3647},
  year={1987},
  publisher={The Physical Society of Japan},
  doi={10.1143/JPSJ.56.3635}
}

@article{garst2017collective,
  title={Collective spin excitations of helices and magnetic skyrmions: review and perspectives of magnonics in non-centrosymmetric magnets},
  author={Garst, Markus and Waizner, Johannes and Grundler, Dirk},
  journal={J. Phys. D: Appl. Phys.},
  volume={50},
  number={29},
  pages={293002},
  year={2017},
  publisher={IOP Publishing},
  doi={10.1088/1361-6463/aa7573}
}

@article{park2020momentum,
  title={{Momentum-dependent magnon lifetime in the metallic noncollinear triangular antiferromagnet CrB$_2$}},
  author={Park, Pyeongjae and Park, Kisoo and Kim, Taehun and Kousaka, Yusuke and Lee, Ki Hoon and Perring, T. G. and Jeong, Jaehong and Stuhr, Uwe and Akimitsu, Jun and Kenzelmann, Michel and others},
  journal={Phys. Rev. Lett.},
  volume={125},
  number={2},
  pages={027202},
  year={2020},
  publisher={APS},
  doi={10.1103/PhysRevLett.125.027202}
}

@article{bannenberg2018magnetization,
  title={{Magnetization and ac susceptibility study of the cubic chiral magnet Mn$_{1-x}$Fe$_x$Si}},
  author={Bannenberg, Lars J and Weber, Frank and Lefering, A. J. E. and Wolf, Thomas and Pappas, Catharine},
  journal={Phys. Rev. B},
  volume={98},
  number={18},
  pages={184430},
  year={2018},
  publisher={APS},
  doi={10.1103/PhysRevB.98.184430}
}

@article{CNCS1,
   author = {G. Ehlers and A. Podlesnyak and J. L. Niedziela and E. B. Iverson and P. E. Sokol},
   title = {The new cold neutron chopper spectrometer at the Spallation Neutron Source: design and performance},
   year = {2011},
   journal = {Rev. Sci. Instrum.},
   volume = {82},
   pages = {085108},
   doi={10.1063/1.3626935}
}

@article{CNCS2,
   author = {G. Ehlers and A. Podlesnyak and A. I. Kolesnikov},
   title = {{The cold neutron chopper spectrometer at the Spallation Neutron Source - A review of the first 8 years of operation}},
   year = {2016},
   journal = {Rev. Sci. Instrum.},
   volume = {87},
   pages = {093902},
   doi={10.1063/1.4962024}
}

@article{sukhanov2019giant,
  title={{Giant enhancement of the skyrmion stability in a chemically strained helimagnet}},
  author={Sukhanov, AS and Vir, Praveen and Heinemann, A and Nikitin, SE and Kriegner, D and Borrmann, H and Shekhar, C and Felser, C and Inosov, DS},
  journal={Phys. Rev. B},
  volume={100},
  number={18},
  pages={180403},
  year={2019},
  publisher={APS},
  doi={10.1103/PhysRevB.100.180403}
}

@article{mishra2023quantum,
  title={{Quantum Griffiths phase in disordered Mn$_{1-x}$Fe$_x$Si}},
  author={Mishra, Ashish Kumar and Samatham, S Shanmukharao and Telling, Mark TF and Hillier, AD and Lees, Martin R and Suresh, KG and Ganesan, V},
  journal={Phys. Rev. B},
  volume={107},
  number={10},
  pages={L100405},
  year={2023},
  publisher={APS},
  doi={10.1103/PhysRevB.107.L100405}
}

@article{demishev2016magnetization,
  title={{Magnetization of Mn$_{1-x}$Fe$_x$Si in high magnetic fields up to 50~T: Possible evidence of a field-induced Griffiths phase}},
  author={S. V. Demishev and A. N. Samarin and J. Huang and V. V. Glushkov and I. I. Lobanova and N. E. Sluchanko and N. M. Chubova and V. A. Dyadkin and S. V. Grigoriev and M. Yu. Kagan and J. Vanacken and V.V. Moshchalkov},
  journal={JETP letters},
  volume={104},
  number={2},
  pages={116--123},
  year={2016},
  publisher={Springer},
  doi={10.1134/S0021364016140022}
}

@article{dahlbom2025sunny,
  title={{Sunny. jl: A Julia package for spin dynamics}},
  author={D. Dahlbom and H. Zhang and C. Miles and S. Quinn and A. Niraula and B. Thipe and M. Wilson and S. Matin and H. Mankad and S. Hahn and D. Pajerowski and S. Johnston and Z. Wang and H. Lane and Y. W. Li and X. Bai and M. Mourigal and C. D. Batista and K. Barros},
  journal={J. Open Source Softw.},
  volume={10},
  number={116},
  pages={8138},
  year={2025},
  publisher = {The Open Journal},
  doi={10.21105/joss.08138}
}

@article{Moriya1960,
  author  = {Moriya, T{\^o}ru},
  title   = {Anisotropic Superexchange Interaction and Weak Ferromagnetism},
  journal = {Phys. Rev.},
  volume  = {120},
  pages   = {91--98},
  year    = {1960},
  doi     = {10.1103/PhysRev.120.91}
}

@article{Mantid,
  title = {{Mantid -- Data analysis and visualization package for neutron scattering and $\mu$SR experiments}},
  author = {O. Arnold and J. C. Bilheux and J. M. Borreguero and A. Buts and S. I. Campbell and L. Chapon and M. Doucet and N. Draper and R. F. Leal and M. A. Gigg and others},
  journal = {Nucl. Instrum. Methods Phys. Res. Sect. A},
  volume = {764},
  pages = {156},
  year = {2014},
  doi={10.1016/j.nima.2014.07.029}
}

@misc{SM,
  howpublished = {See Supplemental Material (SM) for details LSWT calculations, resolution convolution, which includes Refs.~\cite{Moriya1960, dahlbom2025sunny, maleyev2006cubic}}
}


\clearpage
\definecolor{darkblue}{rgb}{0, 0, 0.8}
\graphicspath{{Figures/}}
\renewcommand{\thefigure}{S\arabic{figure}}
\renewcommand{\thetable}{S\arabic{table}}
\renewcommand\theequation{S\arabic{equation}}
\renewcommand\thesection{S\arabic{section}}
\newcommand{\overbar}[1]{\mkern 1.5mu\overline{\mkern-1.5mu#1\mkern-1.5mu}\mkern 1.5mu}
\newcommand{\mb}{$\mu_{\rm B}$}

\onecolumngrid
\begin{center}
{\Large\bfseries Supplemental Material}\\
{\large Anisotropy of spin waves in the field-polarized phase of Fe-doped MnSi}\\
{\normalsize I.~N.~Khoroshiy, A.~Podlesnyak, D.~Menzel, M.~C.~Rahn, D.~S.~Inosov, A.~S.~Sukhanov, and S.~E.~Nikitin}
\end{center}
\twocolumngrid

\section{S1. Effect of dipolar interaction and anisotropic exchange}
\begin{figure}[tb]

\center{\includegraphics[width=.9\linewidth]{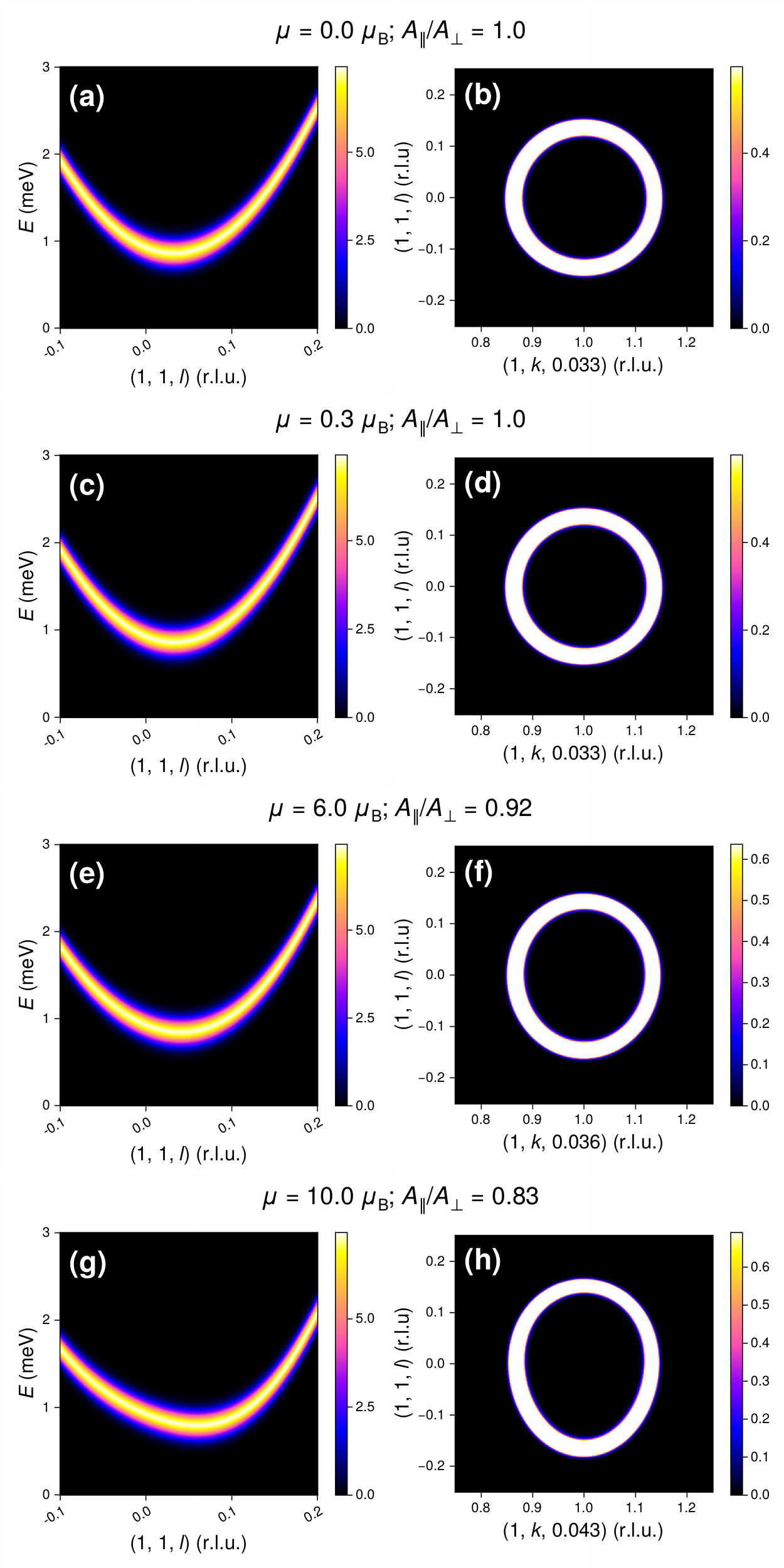}}
  \caption{~Calculated spin-wave spectra including the dipole-dipole interaction. The strength of the dipolar term was artificially tuned to make its effect more visible. This is effectively equivalent to changing the local magnetic moment, whose values are indicated above the corresponding panels. Panels (a,c,e,g) show the spectra along the $(1,1,l)$ direction, while panels (b,d,f,h) show the corresponding constant-energy slices.
  }
  \label{fig_dipole}
  \vspace{-12pt}
\end{figure}

\begin{figure}[tb]
\center{\includegraphics[width=.9\linewidth]{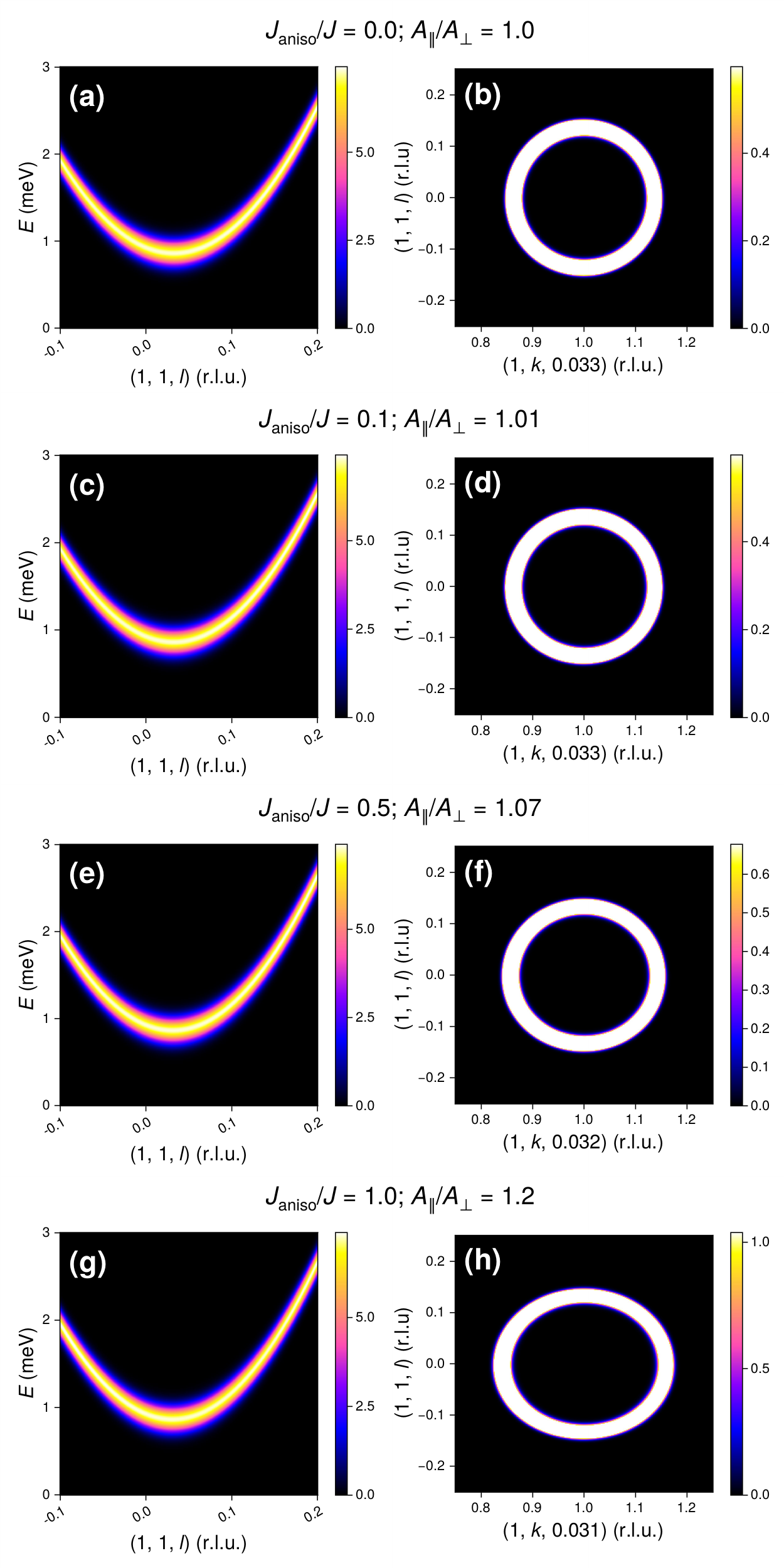}}
  \caption{~Calculated spin-wave spectra including the anisotropic exchange interaction. The values of $J_{\rm aniso}/J$ are indicated above the corresponding panels. Panels (a,c,e,g) show the spectra along the $(1,1,l)$ direction, while panels (b,d,f,h) show the corresponding constant-energy slices.
  }
  \label{fig_anis}
  \vspace{-12pt}
\end{figure}

In this section, we discuss the effects of anisotropic exchange and dipolar interactions on the anisotropy of the spin-wave stiffness in the field-polarized phase of Mn$_{0.9}$Fe$_{0.1}$Si. We show that, although both interactions can in principle produce a finite anisotropy, their contributions in the actual material are expected to be negligible.
We consider the following spin Hamiltonian:

\begin{align}
    \mathcal{H} &= \sum_{i, j} \mathbf{S}_i \hat{J} \mathbf{S}_i + g\mu_{\rm B}\sum_{i} \mathbf{B}\cdot\ \mathbf{S}_i + E_{\rm dipole},\\ 
    \hat{J} &=   \left[
\begin{matrix}
J+J_{\rm aniso}  & D  & -D\\
-D & J-J_{\rm aniso}  & D \\
D  & -D & J
\end{matrix}
\right]
\end{align}

where $J$, $D$, and $J_{\rm aniso}$ denote the isotropic exchange, DMI, and anisotropic exchange interactions, respectively, while $E_{\rm dipole}$ represents the contribution from dipolar interactions.
We performed spin-dynamics calculations within LSWT using the Sunny software package~\cite{dahlbom2025sunny}. We investigated the effects of the dipolar interaction and anisotropic exchange separately, with the corresponding results shown in Fig.~\ref{fig_dipole} and Fig.~\ref{fig_anis}, respectively. For each parameter set, we present the energy spectrum along the field direction in the left panels and a constant-energy slice in the $(1,k,l)$ plane, taken slightly above the dispersion minimum, in the right panels. This slice is used to quantify the dispersion anisotropy through the stiffness ratio $A_{\|}/A_{\perp}$.

As a first step, we calculate the spin-wave dispersion for the model without dipolar interactions or anisotropic exchange. The resulting spectrum exhibits a parabolic dispersion with perfectly circular, isotropic constant-energy contours, characterized by $A_{\|}/A_{\perp} = 1$ [Fig.~\ref{fig_dipole}(a,b)].

\begin{figure}[b]
\center{\includegraphics[width=1\linewidth]{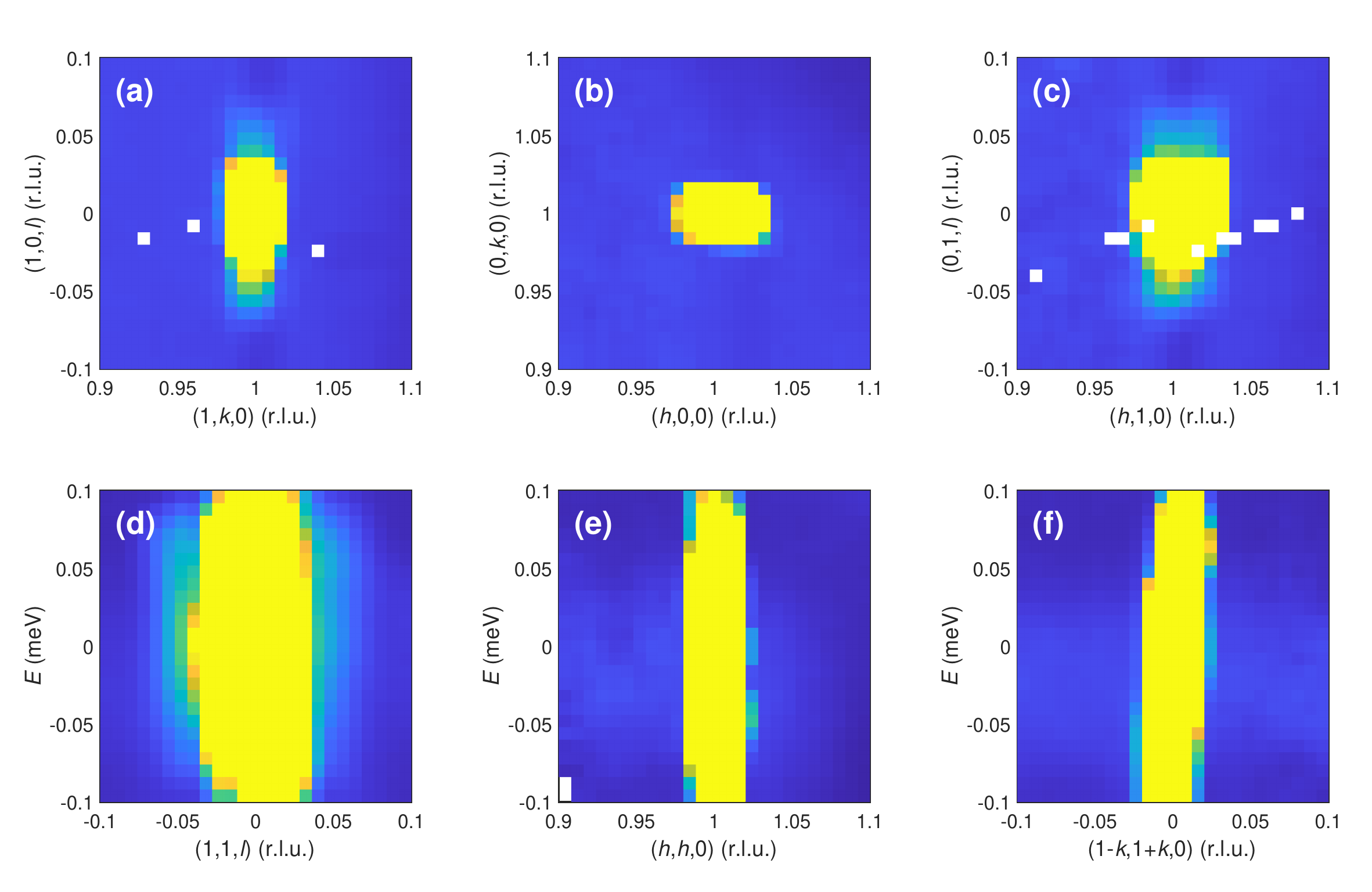}}
  \caption{~
    Two-dimensional slices through the $(1,1,0)$ nuclear Bragg peak. 
  }
  \label{fig_slices}
  \vspace{-12pt}
\end{figure}

Next, we include the dipole-dipole interaction using $m_{\rm Mn} = 0.3~\mu_{\rm B}$ [Fig.~\ref{fig_dipole}(c,d)]. This interaction produces no visible effect on the spectrum. To estimate how large the Mn magnetic moment would need to be to produce a measurable effect, we artificially increased the dipolar energy. We find that a pronounced anisotropy of a few percent emerges only for magnetic moments of several~$\mu_{\rm B}$, far exceeding the experimentally observed value.

\begin{figure}[tb]
\center{\includegraphics[width=1\linewidth]{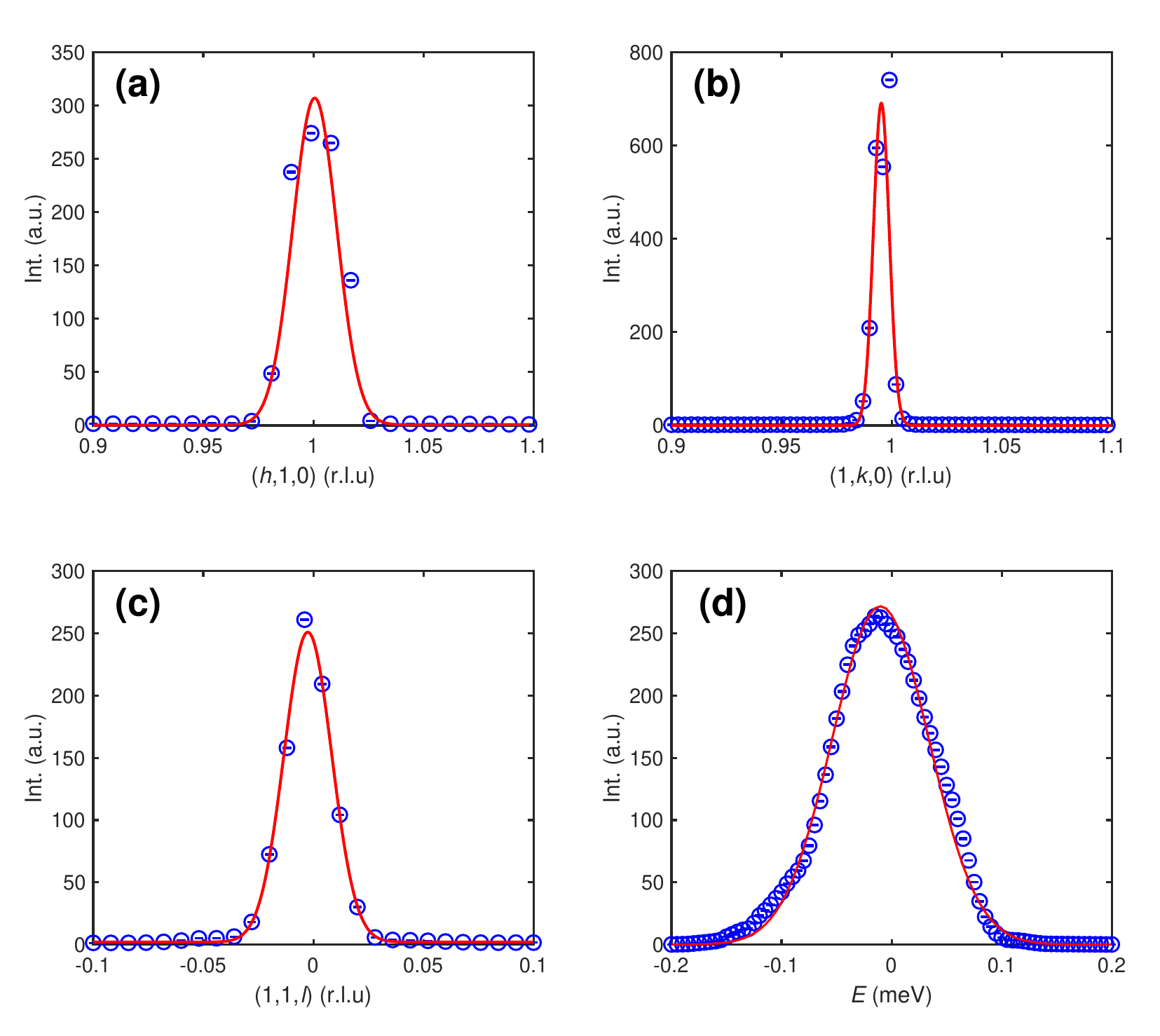}}
  \caption{~ One dimensional cuts through the $(1,1,0)$ Bragg peak along several $\mathbf{Q}$-directions and along energy axis fitted with Gaussian function.
  }
  \label{fig_fits}
  \vspace{-12pt}
\end{figure}

\begin{figure}[b]
\center{\includegraphics[width=1\linewidth]{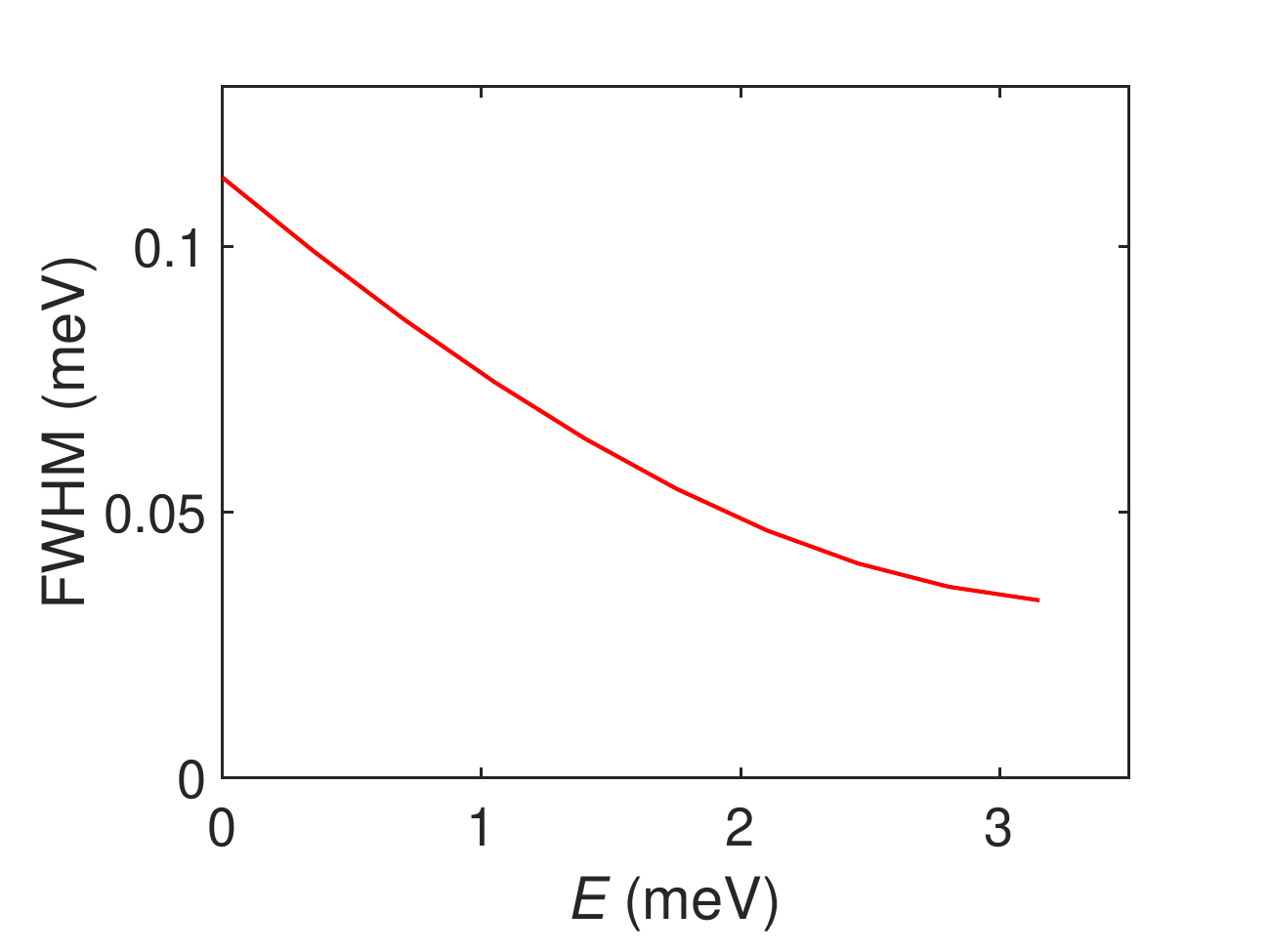}}
  \caption{~Energy dependence of the energy resolution calculated for CNCS with $E_{\rm i} = 3.32$~meV. 
  }
  \label{fig_resolution}
  \vspace{-12pt}
\end{figure}

To investigate the effect of exchange anisotropy, we performed a similar analysis. We calculated the spectra for different values of $J_{\rm aniso}$, with the results summarized in Fig.~\ref{fig_anis}. We find that, even for a sizable anisotropy of $J_{\rm aniso}/J = 0.1$, the spin-wave stiffness ratio deviates from unity by only 0.01, whereas changes on the order of tens of percent are obtained only for $J_{\rm aniso}/J > 0.5$. Such a large anisotropic exchange appears unrealistic for MnSi. Within the standard weak-spin-orbit-coupling theory of cubic helimagnets, the Dzyaloshinskii--Moriya interaction is linear in the spin-orbit-coupling strength, whereas symmetric anisotropic exchange arises at second order, leading to the hierarchy $J \gg D > J_{\rm aniso}$~\cite{Moriya1960, maleyev2006cubic}.

We therefore conclude that, although dipolar interactions and anisotropic exchange can in principle produce anisotropy in the spin-wave stiffness, their energy scales are too small to account for the effect observed in Mn$_{0.9}$Fe$_{0.1}$Si.

\section{S2. Resolution effect}

To account for the effects of energy and momentum resolution, we use an empirical local resolution model based on the measured shape of the nuclear $(1,1,0)$ Bragg reflection. This approach provides a reliable estimate of the instrumental momentum resolution when the sample mosaicity is small. It is particularly suitable for the present experiment because all measurements were performed within a limited $\mathbf{Q}$ range close to the $(1,1,0)$ reflection, where the resolution function determined from the Bragg peak remains a good local approximation.

In Fig.~\ref{fig_slices} we show several slices through the (1,1,0) Bragg peak and the peak is not significantly tilted any plane, meaning that we can make simplification by disentangling the 4D resolution ellipsoid as:
\begin{align}
\mathcal{R}(\mathbf{Q, \omega}) = \mathcal{R}(h)\mathcal{R}(k)\mathcal{R}(l)\mathcal{R}(\omega).    
\end{align}

The momentum-resolution widths were obtained by fitting cuts through the $(1,1,0)$ reflection along the corresponding directions with Gaussian functions, as shown in Fig.~\ref{fig_fits}. The resulting widths are $\Delta h=0.0325$~\AA$^{-1}$, $\Delta k=0.0117$~\AA$^{-1}$, and $\Delta l=0.0352$~\AA$^{-1}$. Here, $\Delta h$, $\Delta k$, and $\Delta l$ denote FWHM along the respective directions. The energy-resolution function $\mathcal{R}(\omega)$ was obtained from the standard energy-dependent resolution calculations for CNCS and is shown in Fig.~\ref{fig_resolution}.

The spin-wave spectra were then fitted using a two-dimensional momentum-energy convolution. For example, for cuts along $(1,1,l)$, the calculated scattering function was convoluted with $\mathcal{R}(l)\mathcal{R}(\omega)$. The corresponding resolution components were used for cuts along the other momentum directions.
The finite integration window perpendicular to the cut direction partially accounts for the transverse momentum resolution. We also verified that moderate variations of this integration window have only a minor effect on the fitted spin-wave stiffnesses. 

The inclusion of the resolution convolution likewise does not qualitatively alter the extracted dispersions or the stiffness anisotropy. In particular, although the vertical momentum resolution is broader than the in-plane resolution, the corresponding broadening is too small to account for the observed difference between $A_{\parallel}$ and $A_{\perp}$.


\end{document}